\documentclass[conference]{IEEEtran}
\IEEEoverridecommandlockouts
\usepackage{cite}
\usepackage{amsmath,amssymb,amsfonts}
\usepackage{algorithmic}
\usepackage{graphicx}
\usepackage{textcomp}
\usepackage{xcolor}
\usepackage{dsfont}
\usepackage{subcaption}
\usepackage{url}
\def\BibTeX{{\rm B\kern-.05em{\sc i\kern-.025em b}\kern-.08em
    T\kern-.1667em\lower.7ex\hbox{E}\kern-.125emX}}

\newcommand{\rev}[1]{#1}

\newcommand{\sy}[1]{\textit{#1}}

\usepackage{listings}
\usepackage{qsharp}
\usepackage[capitalise]{cleveref}
\usepackage{braket}
\usepackage{tikz}
\usepackage[altpo]{backnaur}

\begin{document}

\title{QParallel: Explicit Parallelism for Programming Quantum Computers}

\author{\IEEEauthorblockN{Thomas H\"aner\IEEEauthorrefmark{1}, Vadym Kliuchnikov\IEEEauthorrefmark{2}, Martin Roetteler\IEEEauthorrefmark{2}, Mathias Soeken\IEEEauthorrefmark{1} and Alexander Vaschillo\IEEEauthorrefmark{2}}
\IEEEauthorblockA{\IEEEauthorrefmark{1}Microsoft Quantum, Switzerland}
\IEEEauthorblockA{\IEEEauthorrefmark{2}Microsoft Quantum, USA}}


\maketitle



\begin{abstract}
  We present a language extension for parallel quantum programming to (1) remove ambiguities concerning parallelism in current quantum programming languages and (2) facilitate space-time tradeoff investigations in quantum computing. While the focus of similar libraries in the domain of classical computing (OpenMP, OpenACC, etc.) is to divide a computation into multiple threads, the main goal of QParallel is to keep the compiler and the runtime system from introducing parallelism-inhibiting dependencies, e.g., through reuse of qubits in automatic qubit management. We describe the syntax and semantics of the proposed language extension, implement a prototype based on Q\#, and present several examples and use cases to illustrate its performance benefits. Moreover, we introduce a tool that guides programmers in the placement of parallel regions by identifying the subroutines that profit most from parallelization, which is especially useful if the programmer's knowledge of the source code is limited. Support for QParallel can be added to any multithreading library and language extension, including OpenMP and OpenACC.
\end{abstract}

\begin{IEEEkeywords}
quantum computing
parallel quantum computing
space-time tradeoffs
\end{IEEEkeywords}


\section{Introduction}

Quantum computers promise exponential speedups for certain computational tasks, including problems in chemistry~\cite{mcardle2020quantum} and cryptography~\cite{shor1994algorithms}. 
\rev{This applications require the use of thousands of (error-corrected) qubits and millions of gates.}
Multiple software frameworks and programming languages for quantum computing have emerged~\cite{svore2018q,bichsel2020silq,steiger2018projectq,aleksandrowicz2019qiskit,green2013quipper,javadiabhari2015scaffcc,paykin2017qwire} to assess whether such speedups translate to an advantage of quantum computers over classical supercomputers \textit{in practice}.

These frameworks allow users to implement, test, and debug quantum algorithms, before using the resulting implementation to obtain estimates, e.g., for the number of quantum bits (qubits) or operations required by the algorithm.
Such resource estimates are then used to focus algorithmic optimizations on computational bottlenecks.

The compiler must make several choices when mapping the high-level quantum program to the instruction set architecture (ISA) of the target device.  Examples are which quantum error correction (QEC) procedure to employ and how many magic-state factories to use.
In particular, these choices result in space-time tradeoffs, since (1)~a longer computation requires a larger-distance QEC code and therefore a larger physical-per-logical qubit ratio, and (2)~using more factories (to reduce runtime) limits the space that is left on the quantum chip to place algorithmic qubits.
Moreover, since the rate at which magic states are consumed is not constant throughout the execution of a quantum program \textit{a priori}, additional space-time optimizations are necessary to ensure proper usage of available resources.

For example, while peaks in magic state consumption rate may be handled by magic state buffers, the buffer size required may severely reduce the number of available algorithmic qubits which, in turn, may prohibit other optimizations.
In such cases, it would thus be preferable to modify the quantum program such that the magic state consumption rate is (closer to) constant.  \rev{Given these constraints, automatic parallelization is a hard task and convenient mechanism for manual control are preferable.  This has also been observed in classical computing which motivated solutions such as OpenMP~\cite{dagum1998openmp}.}

Quantum programming languages should therefore make space-time tradeoffs as straightforward as possible.
\rev{In addition, they have to work for applications that require thousands of qubits and millions of gates.
This is a challenge for frameworks that first create a data-structure describing a full quantum circuit and run various 
optimizations on such data structure.}
To this end, we propose an extension for quantum programming languages that gives programmers fine-grained control over which parts of the quantum program run in parallel \rev{and works for large scale applications.}

\subsubsection*{Contributions} We introduce a language extension \rev{akin to OpenMP} that allows quantum programmers to investigate various space-time tradeoffs more efficiently through explicit parallelization.
We implement our proposal as a Q\#~\cite{svore2018q} preprocessor and present several algorithmic benchmarks to illustrate the benefits of our language extension. Specifically, our contributions are as follows.

\begin{itemize}
    \item We introduce the \sy{QParallel} language extension for explicitly-parallel quantum programming. \sy{QParallel} allows programmers to remove ambiguities concerning parallelism in current quantum programming languages (see related work below).
    \item We provide a detailed description of the syntax and semantics of our proposed language extension.
    \item We implement a prototype of \sy{QParallel} in Q\# to allow programmers to define parallel regions \textit{explicitly}
     based on a language feature proposal~\cite{qsproposal}.
    \item We present several examples and use cases that leverage our language extension to achieve improvements in space-time volume.
    \item We show how to improve the integration of our language extension with the quantum software stack using a tool that finds and visualizes critical paths in the quantum program. This is especially valuable when the programmer adding explicit parallelism has limited knowledge of the codebase.
\end{itemize}

\subsubsection*{Related work}
Existing high-level quantum programming languages and frameworks~\cite{svore2018q,bichsel2020silq,steiger2018projectq,green2013quipper,javadiabhari2015scaffcc} have been designed under the assumption that parallelism is \textit{implicit}. This leads to
\begin{enumerate}
    \item unpredictable performance and space-time requirements, and
    \item unexpected interaction with features of the programming language such as quantum memory management.
\end{enumerate}
Moreover, this assumption makes it difficult for programmers to influence the parallelization strategy. We discuss these points in more detail in the next section.
Our work addresses these issues by introducing a way for programmers to be \textit{explicit} about which sections of a quantum program should be executed in parallel.

Our language extension is closely related to classical libraries and language extensions for multithreading such as OpenMP~\cite{dagum1998openmp}, OpenACC~\cite{openacc}, and Thread Building Blocks~\cite{inteltbb}. However, our extension is targeted at \textit{quantum} programming languages and thus addresses several quantum-specific issues that arise in this context, including the interaction of parallel constructs with quantum resource management and the controlled execution of quantum subroutines.

Finally, recent work has introduced QMPI, an extension of MPI supporting communication of quantum data, together with a performance model to develop and analyze distributed quantum programs~\cite{qmpi21}. Similarly, we show that some of the existing work in classical high-performance computing can be leveraged and adapted to solve problems in quantum computing.

\section{Parallelism and Programming Languages}

In classical programming languages such as C++, loops are sequential by default: No programmer would expect the following code snippet (in C++) to run loop iterations in parallel.
\begin{lstlisting}[language=c++]
for (auto b : bits) {
  f(b);
}
\end{lstlisting}
In contrast, for a very similar code fragment written in a quantum programming language such as Q\#, all loop iterations may be executed in parallel on a quantum computer:
\begin{lstlisting}[language=qsharp]
for (q in qubits) {
  H(q);
}
\end{lstlisting}
While this may seem surprising, we believe that the reason for this implicit assumption has its origin in so-called quantum circuit diagrams (see below), which are often used to describe simple quantum algorithms and which are inherently parallel. In the evolution from quantum circuits to high-level programming languages, this implicit assumption has remained intact, resulting in discord between classical and quantum programming.

\subsection{From quantum circuits to code}

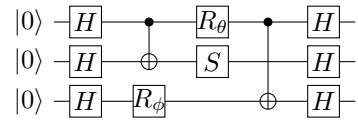
\begin{figure}[t]
\centering
\begin{tikzpicture}[scale=1.000000,x=1pt,y=1pt]
\filldraw[color=white] (0.000000, -7.500000) rectangle (114.000000, 37.500000);
\draw[color=black] (0.000000,30.000000) -- (114.000000,30.000000);
\draw[color=black] (0.000000,30.000000) node[left] {$|0\rangle$};
\draw[color=black] (0.000000,15.000000) -- (114.000000,15.000000);
\draw[color=black] (0.000000,15.000000) node[left] {$|0\rangle$};
\draw[color=black] (0.000000,0.000000) -- (114.000000,0.000000);
\draw[color=black] (0.000000,0.000000) node[left] {$|0\rangle$};
\begin{scope}
\draw[fill=white] (12.000000, 30.000000) +(-45.000000:8.485281pt and 8.485281pt) -- +(45.000000:8.485281pt and 8.485281pt) -- +(135.000000:8.485281pt and 8.485281pt) -- +(225.000000:8.485281pt and 8.485281pt) -- cycle;
\clip (12.000000, 30.000000) +(-45.000000:8.485281pt and 8.485281pt) -- +(45.000000:8.485281pt and 8.485281pt) -- +(135.000000:8.485281pt and 8.485281pt) -- +(225.000000:8.485281pt and 8.485281pt) -- cycle;
\draw (12.000000, 30.000000) node {$H$};
\end{scope}
\begin{scope}
\draw[fill=white] (12.000000, 15.000000) +(-45.000000:8.485281pt and 8.485281pt) -- +(45.000000:8.485281pt and 8.485281pt) -- +(135.000000:8.485281pt and 8.485281pt) -- +(225.000000:8.485281pt and 8.485281pt) -- cycle;
\clip (12.000000, 15.000000) +(-45.000000:8.485281pt and 8.485281pt) -- +(45.000000:8.485281pt and 8.485281pt) -- +(135.000000:8.485281pt and 8.485281pt) -- +(225.000000:8.485281pt and 8.485281pt) -- cycle;
\draw (12.000000, 15.000000) node {$H$};
\end{scope}
\begin{scope}
\draw[fill=white] (12.000000, -0.000000) +(-45.000000:8.485281pt and 8.485281pt) -- +(45.000000:8.485281pt and 8.485281pt) -- +(135.000000:8.485281pt and 8.485281pt) -- +(225.000000:8.485281pt and 8.485281pt) -- cycle;
\clip (12.000000, -0.000000) +(-45.000000:8.485281pt and 8.485281pt) -- +(45.000000:8.485281pt and 8.485281pt) -- +(135.000000:8.485281pt and 8.485281pt) -- +(225.000000:8.485281pt and 8.485281pt) -- cycle;
\draw (12.000000, -0.000000) node {$H$};
\end{scope}
\draw (36.000000,30.000000) -- (36.000000,15.000000);
\begin{scope}
\draw[fill=white] (36.000000, 15.000000) circle(3.000000pt);
\clip (36.000000, 15.000000) circle(3.000000pt);
\draw (33.000000, 15.000000) -- (39.000000, 15.000000);
\draw (36.000000, 12.000000) -- (36.000000, 18.000000);
\end{scope}
\filldraw (36.000000, 30.000000) circle(1.500000pt);
\begin{scope}
\draw[fill=white] (36.000000, -0.000000) +(-45.000000:8.485281pt and 8.485281pt) -- +(45.000000:8.485281pt and 8.485281pt) -- +(135.000000:8.485281pt and 8.485281pt) -- +(225.000000:8.485281pt and 8.485281pt) -- cycle;
\clip (36.000000, -0.000000) +(-45.000000:8.485281pt and 8.485281pt) -- +(45.000000:8.485281pt and 8.485281pt) -- +(135.000000:8.485281pt and 8.485281pt) -- +(225.000000:8.485281pt and 8.485281pt) -- cycle;
\draw (36.000000, -0.000000) node {$R_\phi$};
\end{scope}
\begin{scope}
\draw[fill=white] (60.000000, 30.000000) +(-45.000000:8.485281pt and 8.485281pt) -- +(45.000000:8.485281pt and 8.485281pt) -- +(135.000000:8.485281pt and 8.485281pt) -- +(225.000000:8.485281pt and 8.485281pt) -- cycle;
\clip (60.000000, 30.000000) +(-45.000000:8.485281pt and 8.485281pt) -- +(45.000000:8.485281pt and 8.485281pt) -- +(135.000000:8.485281pt and 8.485281pt) -- +(225.000000:8.485281pt and 8.485281pt) -- cycle;
\draw (60.000000, 30.000000) node {$R_\theta$};
\end{scope}
\begin{scope}
\draw[fill=white] (60.000000, 15.000000) +(-45.000000:8.485281pt and 8.485281pt) -- +(45.000000:8.485281pt and 8.485281pt) -- +(135.000000:8.485281pt and 8.485281pt) -- +(225.000000:8.485281pt and 8.485281pt) -- cycle;
\clip (60.000000, 15.000000) +(-45.000000:8.485281pt and 8.485281pt) -- +(45.000000:8.485281pt and 8.485281pt) -- +(135.000000:8.485281pt and 8.485281pt) -- +(225.000000:8.485281pt and 8.485281pt) -- cycle;
\draw (60.000000, 15.000000) node {$S$};
\end{scope}
\draw (81.000000,30.000000) -- (81.000000,0.000000);
\begin{scope}
\draw[fill=white] (81.000000, 0.000000) circle(3.000000pt);
\clip (81.000000, 0.000000) circle(3.000000pt);
\draw (78.000000, 0.000000) -- (84.000000, 0.000000);
\draw (81.000000, -3.000000) -- (81.000000, 3.000000);
\end{scope}
\filldraw (81.000000, 30.000000) circle(1.500000pt);
\begin{scope}
\draw[fill=white] (102.000000, 30.000000) +(-45.000000:8.485281pt and 8.485281pt) -- +(45.000000:8.485281pt and 8.485281pt) -- +(135.000000:8.485281pt and 8.485281pt) -- +(225.000000:8.485281pt and 8.485281pt) -- cycle;
\clip (102.000000, 30.000000) +(-45.000000:8.485281pt and 8.485281pt) -- +(45.000000:8.485281pt and 8.485281pt) -- +(135.000000:8.485281pt and 8.485281pt) -- +(225.000000:8.485281pt and 8.485281pt) -- cycle;
\draw (102.000000, 30.000000) node {$H$};
\end{scope}
\begin{scope}
\draw[fill=white] (102.000000, 15.000000) +(-45.000000:8.485281pt and 8.485281pt) -- +(45.000000:8.485281pt and 8.485281pt) -- +(135.000000:8.485281pt and 8.485281pt) -- +(225.000000:8.485281pt and 8.485281pt) -- cycle;
\clip (102.000000, 15.000000) +(-45.000000:8.485281pt and 8.485281pt) -- +(45.000000:8.485281pt and 8.485281pt) -- +(135.000000:8.485281pt and 8.485281pt) -- +(225.000000:8.485281pt and 8.485281pt) -- cycle;
\draw (102.000000, 15.000000) node {$H$};
\end{scope}
\begin{scope}
\draw[fill=white] (102.000000, -0.000000) +(-45.000000:8.485281pt and 8.485281pt) -- +(45.000000:8.485281pt and 8.485281pt) -- +(135.000000:8.485281pt and 8.485281pt) -- +(225.000000:8.485281pt and 8.485281pt) -- cycle;
\clip (102.000000, -0.000000) +(-45.000000:8.485281pt and 8.485281pt) -- +(45.000000:8.485281pt and 8.485281pt) -- +(135.000000:8.485281pt and 8.485281pt) -- +(225.000000:8.485281pt and 8.485281pt) -- cycle;
\draw (102.000000, -0.000000) node {$H$};
\end{scope}
\end{tikzpicture}
\caption{Example of a quantum circuit: Qubits are represented by wires with time advancing from left to right. Operations on qubits correspond to boxes and other symbols on the respective wire(s).}
\label{fig:circ}
\end{figure}

In a quantum circuit, qubits are represented by horizontal lines, with time advancing from left to right. Operations on these qubits are depicted as boxes and symbols that are placed on the respective lines, see \cref{fig:circ} for an example. While quantum circuits can be used to describe simple algorithms, their capabilities concerning measurement feedback and other classical control flow are severely limited.

Quantum assembly languages such as OpenQASM were introduced~\cite{cross2017open} to enable expressing simple mixed quantum-classical programs such as quantum teleportation~\cite{bennett1993teleporting}. However, their support for classical control flow is limited to conditional gate applications (e.g., conditional on a measurement outcome). In the meantime, multiple higher-level quantum programming languages have been proposed~\rev{\cite{svore2018q,javadiabhari2015scaffcc,green2013quipper,crossOpenQasm3,mintz2019qcor,bichsel2020silq}} that, among other useful features, support a more general classical control flow.

\subsection{Quantum memory management and parallelism}

One of the main advantages of a high-level quantum programming language is automatic memory management: Instead of having to allocate one array of qubits at the beginning and then passing along the right qubits to every subroutine, memory management allows programmers to allocate and deallocate qubits inside subroutines.

There is, however, a crucial interaction between parallelism and memory management in quantum computing: Depending on the memory management strategy, the compiler (or the run-time system) may introduce artificial dependencies, e.g., among loop iterations that would otherwise be subject to parallelism. As an example, consider the following code snippet:
\begin{lstlisting}[language=qsharp]
for (q in qubits) {
  Op(q);
}
\end{lstlisting}
\sy{Op} is some quantum operation that takes a qubit as input. If \sy{Op} is in the native gate set of the target architecture, for example, then this loop may be executed in parallel.

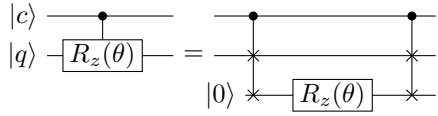
\begin{figure}[t]
\centering
\begin{tikzpicture}[scale=1.000000,x=1pt,y=1pt]
\filldraw[color=white] (0.000000, -7.500000) rectangle (147.000000, 37.500000);
\draw[color=black] (0.000000,30.000000) -- (147.000000,30.000000);
\draw[color=black] (0.000000,30.000000) node[left] {$|c\rangle$};
\draw[color=black] (0.000000,15.000000) -- (147.000000,15.000000);
\draw[color=black] (0.000000,15.000000) node[left] {$|q\rangle$};
\draw[color=black] (75.000000,0.000000) -- (147.000000,0.000000);
\draw[color=black] (75.000000,0.000000) node[left] {$|0\rangle$};
\draw (21.000000,30.000000) -- (21.000000,15.000000);
\begin{scope}
\draw[fill=white] (21.000000, 15.000000) +(-45.000000:21.213203pt and 8.485281pt) -- +(45.000000:21.213203pt and 8.485281pt) -- +(135.000000:21.213203pt and 8.485281pt) -- +(225.000000:21.213203pt and 8.485281pt) -- cycle;
\clip (21.000000, 15.000000) +(-45.000000:21.213203pt and 8.485281pt) -- +(45.000000:21.213203pt and 8.485281pt) -- +(135.000000:21.213203pt and 8.485281pt) -- +(225.000000:21.213203pt and 8.485281pt) -- cycle;
\draw (21.000000, 15.000000) node {$R_z(\theta)$};
\end{scope}
\filldraw (21.000000, 30.000000) circle(1.500000pt);
\draw[fill=white,color=white] (48.000000, 8.000000) rectangle (63.000000, 36.000000);
\draw (55.500000, 15.000000) node {$=$};
\draw (78.000000,30.000000) -- (78.000000,0.000000);
\begin{scope}
\draw (75.878680, 12.878680) -- (80.121320, 17.121320);
\draw (75.878680, 17.121320) -- (80.121320, 12.878680);
\end{scope}
\begin{scope}
\draw (75.878680, -2.121320) -- (80.121320, 2.121320);
\draw (75.878680, 2.121320) -- (80.121320, -2.121320);
\end{scope}
\filldraw (78.000000, 30.000000) circle(1.500000pt);
\begin{scope}
\draw[fill=white] (108.000000, -0.000000) +(-45.000000:21.213203pt and 8.485281pt) -- +(45.000000:21.213203pt and 8.485281pt) -- +(135.000000:21.213203pt and 8.485281pt) -- +(225.000000:21.213203pt and 8.485281pt) -- cycle;
\clip (108.000000, -0.000000) +(-45.000000:21.213203pt and 8.485281pt) -- +(45.000000:21.213203pt and 8.485281pt) -- +(135.000000:21.213203pt and 8.485281pt) -- +(225.000000:21.213203pt and 8.485281pt) -- cycle;
\draw (108.000000, -0.000000) node {$R_z(\theta)$};
\end{scope}
\draw (138.000000,30.000000) -- (138.000000,0.000000);
\begin{scope}
\draw (135.878680, 12.878680) -- (140.121320, 17.121320);
\draw (135.878680, 17.121320) -- (140.121320, 12.878680);
\end{scope}
\begin{scope}
\draw (135.878680, -2.121320) -- (140.121320, 2.121320);
\draw (135.878680, 2.121320) -- (140.121320, -2.121320);
\end{scope}
\filldraw (138.000000, 30.000000) circle(1.500000pt);
\end{tikzpicture}
\caption{Quantum circuit for implementing a controlled $R_z$-rotation. Left: circuit notation for a controlled $R_z$ rotation, right: implementation using one extra qubit in $\ket0$, two Fredkin gates (controlled Swap), and a single-qubit rotation $R_z$.}
\label{fig:crz}
\end{figure}

However, if \sy{Op} temporarily allocates and then deallocates a helper qubit, quantum memory management may decide to reuse the same helper qubit for all iterations of the loop. An example of this is given in \cref{fig:crz} in which on the right implementation of a controlled rotation gate is shown that uses one helper qubit to avoid doubling the number of single-qubit rotations\rev{, which induces a large overhead when implemented fault-tolerantly}.\footnote{\rev{Note that this might not be a preferable circuit transformation in noisy-intermediate scale quantum computers (NISQ).}} Consequently, all potential parallelism is lost due to the introduction of this dependency. We note that the no-cloning theorem prohibits removing this dependency through copying.
Furthermore, an alternative resource management strategy that opts to never reuse resources would result in prohibitively large numbers of qubits and gates.
As a solution to this problem, we thus propose language constructs to explicitly mark parallel regions.

Specifically, using our proposed language extension, the example above could be expressed as
\begin{lstlisting}[language=qsharp]
parallel for (q in qubits) {
  Op(q);
}
\end{lstlisting}
where the \sy{parallel} indicates that managed resources should not be reused across loop iterations.

In addition, we argue that the memory management system should be extended to include the management of so-called \textit{resource states}. For example, automatic management of Fourier states
\[
    \ket F = \frac{1}{\sqrt{N}}\sum_{k=0}^{N-1} e^{ik/N}\ket{k},
\]
where $N=2^n$ with $n$ denoting the number of qubits, can ensure reuse of this resource state when applying multiplexed rotations using an adder, which is cheaper than applying $n$ rotation gates in a fault-tolerant setting~\cite{kitaev2002classical}.

By letting the memory management system handle resource states, it is possible to automatically balance the usage of physical qubit resources for encoding (1) algorithm qubits, (2) magic state factories, and (3) resource states.

\section{Syntax and semantics of QParallel}
\label{sec:syntax}

In this section, we present the syntax and semantics of our proposed language extension. QParallel provides support for parallel sections and loops, in addition to specific clauses that facilitate making existing code parallel.

\subsection{Parallel sections}

A \sy{parallel sections} statement defines a portion of the program that should be run in parallel. The syntax is as follows:
\begin{lstlisting}[style=Qsharp]
parallel sections {
  SectionSequence
}
\end{lstlisting}
The code encapsulated in a \sy{parallel sections} block contains a sequence of section statements. Each section should be run in parallel with other section code blocks, if possible.

\subsection{Section statement}

The section statement marks a section that can be run in parallel with other section blocks in the same \sy{parallel sections} environment.
\begin{lstlisting}[style=Qsharp]
section {
  StatementBlock
}
\end{lstlisting}
A precondition to the usage of this statement is that each section can be executed in parallel. Consequently, different sections must not share any qubits. In particular, to ensure parallel execution, automatic memory management must not reuse temporary qubits that are allocated and deallocated inside a different section of the same parallel sections block.

\subsection{Parallel for statement}

Another case that commonly benefits from parallelization is a loop where each iteration of the loop can be run in parallel. To support this use case, we introduce a \sy{parallel for} statement. Its behavior is identical to that of a \sy{parallel sections} statement, where each iteration of a loop is considered a separate section inside of it. The \sy{parallel for} statement can be used as follows:
\begin{lstlisting}[style=Qsharp]
parallel for i in iterable {
  StatementBlock
}
\end{lstlisting}

\subsection{Fanout clause}

Consider the following for-loop in Q\#:
\begin{lstlisting}[language=qsharp]
for t in targets {
  CNOT(control, t);
}
\end{lstlisting}
This loop cannot be executed in parallel because all \sy{CNOT} operations share the same \sy{control} qubit, declared outside of the loop. As a remedy, we introduce the so-called \sy{fanout} clause, which allows us to rewrite the code above as a parallel for-loop with \sy{fanout}:
\begin{lstlisting}[language=qsharp]
parallel for t in targets fanout(control, 4) {
  CNOT(control, t);
}
\end{lstlisting}
In this example, the control qubit is fanned out to three extra qubits, resulting in a total of 4 control qubits that are shared among loop iterations. This reduces the runtime of the loop by up to $4\times$.

More generally, a \sy{fanout(qubits, n)} clause fans out the specified \sy{qubits} to $n-1$ temporary qubits, so that loop iterations can use one of these ``entangled copies'' and thus remove some or all of the data dependencies. At the end of the loop, an unfanout operation, which is the inverse of the initial fanout operation, will consolidate the effect of all iterations on their respective copies of the fanned-out qubit(s).

In our prototype implementation, the \sy{fanout} clause is currently only supported by the \sy{parallel for} statement but could, in principle, be allowed in other parallel regions.

\subsection{Complete extension syntax}

Here, we detail the syntax of our proposed language extension by providing its Backus-Naur form in \cref{fig:bnf}.


\begin{figure*}
\begin{bnf*}
\bnfprod{parallel-sections} {\sy{parallel sections} \bnfsp \{ \bnfsp \bnfpn{section-sequence} \bnfsp \}} \\
\bnfprod{section-sequence} {\sy{section} \bnfsp \bnfpn{statement-block} \bnfor}\\
& & \sy{section} \bnfsp \bnfpn{statement-block} \bnfsp \bnfpn{section-sequence} \\
\bnfprod{statement-block} {\{ \bnfsp \bnfpn{statement-sequence} \bnfsp \}} \\
\bnfprod{statement-sequence} {\bnfpn{statement} \bnfor \bnfpn{statement} \bnfsp ; \bnfsp \bnfpn{statement-sequence}} \\
\bnfprod{parallel-for} {\sy{parallel} \bnfsp \bnfpn{for-header} \bnfsp \bnfpn{for-body} \bnfor} \\
& & \sy{parallel} \bnfsp \bnfpn{for-header} \bnfsp \bnfpn{fanout} \bnfsp \bnfpn{for-body} \\
\bnfprod{fanout} {\sy{fanout} \bnfsp ( \bnfsp \bnfpn{qubit-array} \bnfsp , \bnfsp \bnfpn{expression} \bnfsp )}
\end{bnf*}
\caption{Backus-Naur form describes the syntax of our proposed language extension.}
\label{fig:bnf}
\end{figure*}

Our extension is compatible with a large set of languages that may have a slightly different syntax for its regular statements. For example, to apply our extension to the Q\# grammar~\cite{qsharpgrammar}, we use a mapping to Q\# nonterminals as follows:
\begin{bnf*}
\bnfprod{for-header} { \sy{for} \bnfsp \bnfpn{forBinding} \bnfsp | \bnfsp \sy{for} \bnfsp ( \bnfsp \bnfpn{forBinding} \bnfsp ) } \\
\bnfprod{for-body} {\bnfpn{scope} } \\
\bnfprod{qubit-array} {\bnfpn{Identifier} } \\
\end{bnf*}

\section{\rev{Extending a quantum programming language with Support for Parallelism}}

\rev{Our approach can be applied to many quantum programming languages. Here we use Q\# as an example to illustrate key implementation details. }
Q\# is a quantum programming language that includes quantum-specific language constructs into a classical imperative programming paradigm~\cite{svore2018q,qsharpdocs}.  These quantum-specific language constructs include qubit allocation and deallocation, inverse and controlled execution of operations, and conjugation control flow.  The latter provides syntax for the common pattern $O_1 \,;\, O_2 \,;\, O_1^\dagger$, i.e., performing some operation $O_1$, followed by another operation $O_2$, and finalized by calling the inverse of $O_1$.  Q\# allows programmers to explicitly state this intent using

\begin{minipage}{\linewidth}
\begin{lstlisting}[style=Qsharp]
within {
  Operation1(...);
} apply {
  Operation2(...);
}
\end{lstlisting}
\end{minipage}

The example in \cref{lst:qsharp-example} illustrates some of these constructs: The operation \sy{ApplyRotations} applies $n$ controlled $R_z$ rotations with angles $\frac{i\pi}{n}$ (for $0 \le i < n$) with two disjoint sets of control and target qubits that are allocated at the beginning of the operation.  The operation \sy{ControlledRz} implements a controlled rotation using a non-controlled $R_z$ rotation together with a helper qubit conjugated by a controlled SWAP operation, see~\cref{fig:crz}.  The helper qubit is allocated locally in this operation, and therefore also deallocated on leaving the operation.

\begin{lstlisting}[float=t,style=Qsharp,caption={This Q\# code applies several controlled $R_z$ rotations in \sy{ApplyRotations}. The controlled rotation operation \sy{ControlledRz} is implemented using a non-controlled rotation and a helper qubit.},label={lst:qsharp-example}]
operation ControlledRz(angle : Double, control : Qubit, target : Qubit) : Unit is Adj {
  use helper = Qubit();

  within {
    Controlled SWAP([control], (helper, target));
  } apply {
    Rz(angle, helper);
  }
}

operation ApplyRotations(n : Int) : Unit {
  use ctls = Qubit[n];
  use tgts = Qubit[n];

  for (i, (ctl, tgt)) in Enumerated(Zipped(ctls, tgts)) {
    ControlledRz(PI() * IntAsDouble(i) / IntAsDouble(n), ctl, tgt);
  }
}
\end{lstlisting}

The Q\# code can be executed against several targets, including a full-state simulator, a resources estimator, or an actual quantum computer.  The target provides information on what to do for \sy{intrinsic} operations such as CNOT gates (used in the implementation of SWAP), $R_z$ gates, and also qubit allocation.

Since qubit allocation is target-dependent, it is unclear from the target-independent Q\# description, whether the controlled $R_z$ rotations can be applied in parallel.  Whereas the control and target qubits are disjoint, the helper qubit in the implementation of \sy{ControlledRz} operation might happen to be the same qubit depending on the qubit allocation strategy of the target.

\sy{QParallel} allows us to avoid such ambiguities.  A \sy{parallel} block indicates that qubit allocation should target maximum parallel execution of quantum operations in this block, whereas outside of \sy{parallel} blocks, qubits are reused whenever possible.  The parallelism in \sy{ApplyRotations} of~\cref{lst:qsharp-example} can be made explicit by merely adding the word \sy{parallel} in front of the for-loop, see \cref{lst:qsharp-example-parallel}.

\begin{lstlisting}[float=t,style=Qsharp,caption={A modified implementation of \sy{ApplyRotations} in \cref{lst:qsharp-example} that uses the proposed \sy{parallel} keyword to indicate parallel execution of quantum operations in the for-loop.},label={lst:qsharp-example-parallel}]
operation ApplyRotations(n : Int) : Unit {
  use ctls = Qubit[n];
  use tgts = Qubit[n];

  parallel for (i, (ctl, tgt)) in Enumerated(Zipped(ctls, tgts)) {
     ControlledRz(PI() * IntAsDouble(i) / IntAsDouble(n), ctl, tgt);
  }
}
\end{lstlisting}

\subsubsection{Implementation Details}
We enable support for our language extension in Q\# through a preprocessing step that maps \sy{QParallel} features to custom instructions that we implement directly in the C\# runtime. In particular, these instructions interact with the quantum memory management to enable or disable qubit reuse in parallel sections.
\rev{Note that our approach is not limited to Q\#, but any quantum programming language that supports configurable quantum memory management through a runtime can be extended in a similar way.}
The preprocessor maps all newly introduced language features to valid Q\# syntax. More specifically, we map these to the special instructions \sy{Parallel}, \sy{Section}, \sy{Fanout}, \sy{Unfanout}, and \sy{GetControlQubits}.

Parallel sections are conjugated by a call to the special instruction \sy{Parallel}:
\[
\frac
{\sy{parallel sections}\bnfsp \bnfpn{section-sequence}}
{\sy{within} \bnfsp \{ \bnfsp \sy{Parallel();} \bnfsp \}
\bnfsp \sy{apply} \bnfsp \bnfpn{section-sequence}}
\]

\noindent Similarly, each section inside is wrapped into calls to the special \sy{Section} instruction:
\[
\frac
{\sy{section} \bnfsp \bnfpn{statement-block}}
{\sy{within} \bnfsp \{ \bnfsp \sy{Section();} \bnfsp \} \bnfsp \sy{apply} \bnfsp \bnfpn{statement-block}}
\]

\noindent The \sy{parallel for} syntax conveniently combines these two:
\[
\frac
{\sy{parallel for} \bnfsp \bnfpn{for-header} \bnfsp \bnfpn{for-body}}
{\begin{aligned}
\sy{within} \bnfsp \{ \bnfsp \sy{Parallel();} \bnfsp \} \bnfsp \sy{apply}
\bnfsp \{ \bnfsp \bnfpn{for-header} \bnfsp \{ \bnfsp \\
\sy{within} \bnfsp \{ \bnfsp \sy{Section();} \bnfsp \} \bnfsp
\sy{apply} \bnfsp \bnfpn{for-body} \bnfsp \} \bnfsp \}
\end{aligned}
}
\]
In addition, a \sy{parallel for} loop can be equipped with the \sy{fanout} keyword. Qubits passed to \sy{fanout} will be fanned out before entering the loop, allowing different loop iterations to use different control qubits and, thus, to be executed in parallel. We refer to \cref{fig:parformapping} for the detailed mapping carried out by our Q\# preprocessor.

\begin{figure*}[t]
\[
\frac
{\sy{parallel for} \bnfsp \bnfpn{for-header} \bnfsp \sy{fanout(} \bnfsp \bnfpn{qubit-array}, \bnfsp \bnfpn{expression} \bnfsp \sy{)} \bnfsp \bnfpn{for-body}}
{%
\begin{array}{l}
\sy{let} \bnfsp \bnfpn{fanout-id} \bnfsp \sy{= Fanout(} \bnfsp \bnfpn{qubit-array} \sy{,} \bnfsp \bnfpn{expression} \bnfsp \sy{);} \\
\sy{within} \bnfsp \{ \bnfsp \sy{Parallel();} \bnfsp \} \bnfsp \sy{apply}
\bnfsp \{ \\
\quad\bnfsp \bnfpn{for-header} \bnfsp \{ \\
\quad\quad\sy{within} \bnfsp \{ \bnfsp \sy{Section();} \bnfsp \} \bnfsp
\sy{apply} \bnfsp \{ \\
\quad\quad\quad\sy{let}\bnfsp\bnfpn{qubit-array}'\bnfsp\sy{= GetControlQubits(}\bnfsp \bnfpn{fanout-id} \bnfsp \sy{);} \\
\quad\quad\quad\bnfpn{for-body}' \bnfsp \\
\quad\quad\} \\
\quad\} \\
\} \\
\sy{Unfanout();}
\end{array}}
\]
\caption{Mapping parallel for loops with a \sy{fanout} clause to Q\#. Here, $\bnfpn{for-body}'$ is obtained by replacing all occurrences of $\bnfpn{qubit-array}$ in $\bnfpn{for-body}$ with $\bnfpn{qubit-array}'$. \sy{GetControlQubits} extracts a copy of the fanned out qubit array by using the fanout-id obtained from the call to \sy{Fanout} (which is handled by the runtime).}
\label{fig:parformapping}
\end{figure*}

All quantum memory is managed by the \sy{QubitManager} from the Q\#-runtime.
\rev{ Qubit manager is the runtime component responsible for allocating and deallocating qubits in Q\#-runtime.
Its standard behaviour is to maintain a pool of qubits. Newly allocated qubits are teaken from the pool and released qubits are returned into the pool.}
The intrinsics that are placed in the code by our preprocessing step change the behavior of the \sy{QubitManager}. Specifically, at the start of each parallel section, the \sy{QubitManager} initializes a new pool of qubits to be allocated from. This pool is reserved exclusively for this section and cannot be accessed from other sections. When qubits are deallocated, they are returned to this reserved pool and can be reallocated again within the same section, but not in any other section. At the end of the parallel block, the reserved pool is returned to the general pool of freely available qubits.
This guarantees that when sections are run in parallel, they do not share newly allocated qubits and, consequently, no artificial data dependencies are introduced that would serialize an otherwise parallel quantum program.

\rev{In conclusion, proposed approach can be easily integrated with any quantum programming language with extensible runtime,
such as Quipper \cite{green2013quipper}, ProjectQ \cite{steiger2018projectq} and QCOR\cite{mintz2019qcor}.
In principle, it is not necessary to modify the language's syntax. 
It is sufficient to extend the language's runtime with additional instructions.
The syntax extension can be viewed as an additional step that helps improve clarity of the source code.
}

\section{Practical Applications}

\begin{figure*}[t]
    \begin{subfigure}[t]{0.5\textwidth}
    \centering
    \resizebox{0.8\linewidth}{!}{\includegraphics{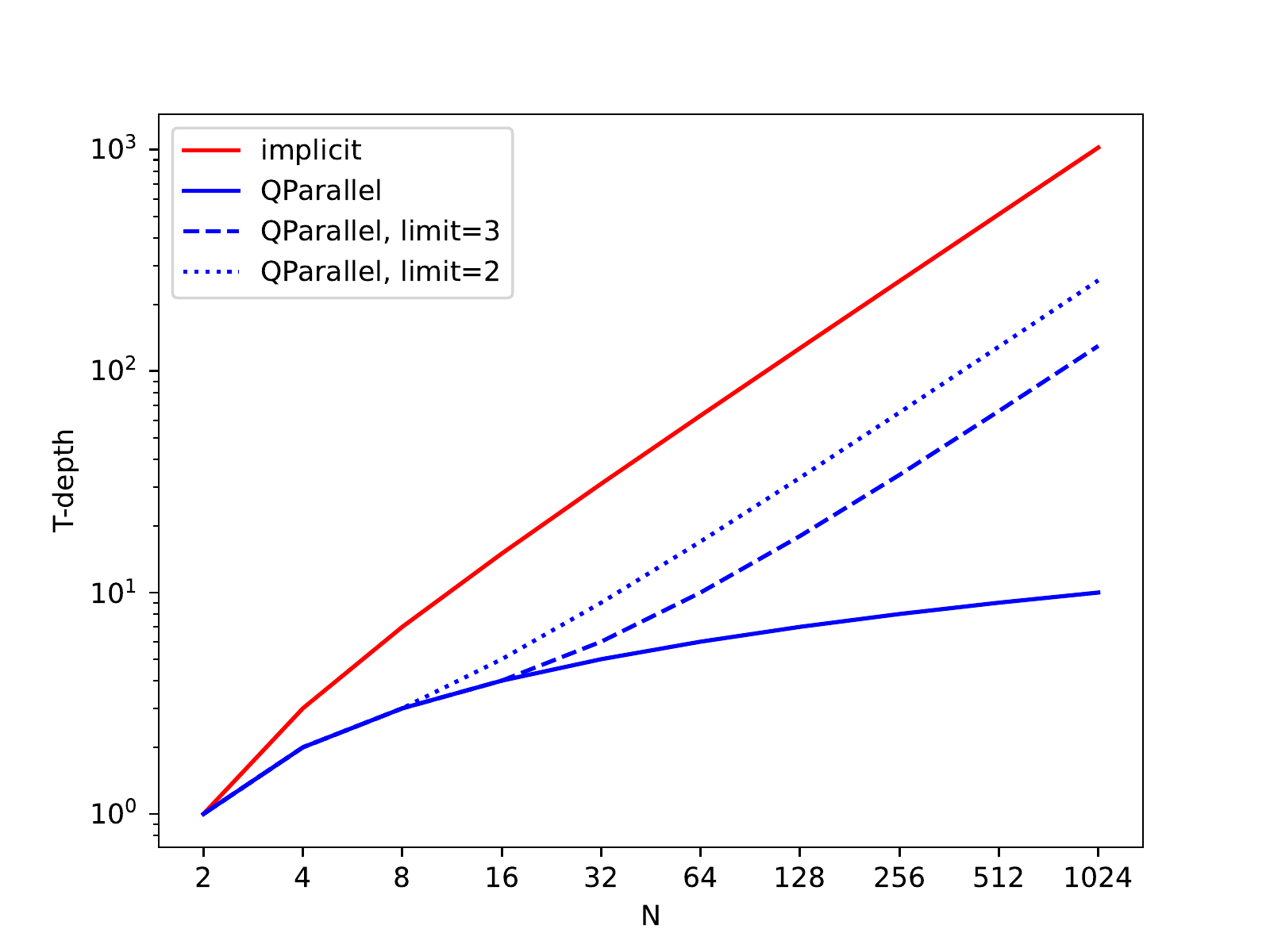}}
    \caption{$T$-depth}
    \end{subfigure}
    \begin{subfigure}[t]{0.5\textwidth}
    \centering
    \resizebox{0.8\linewidth}{!}{\includegraphics{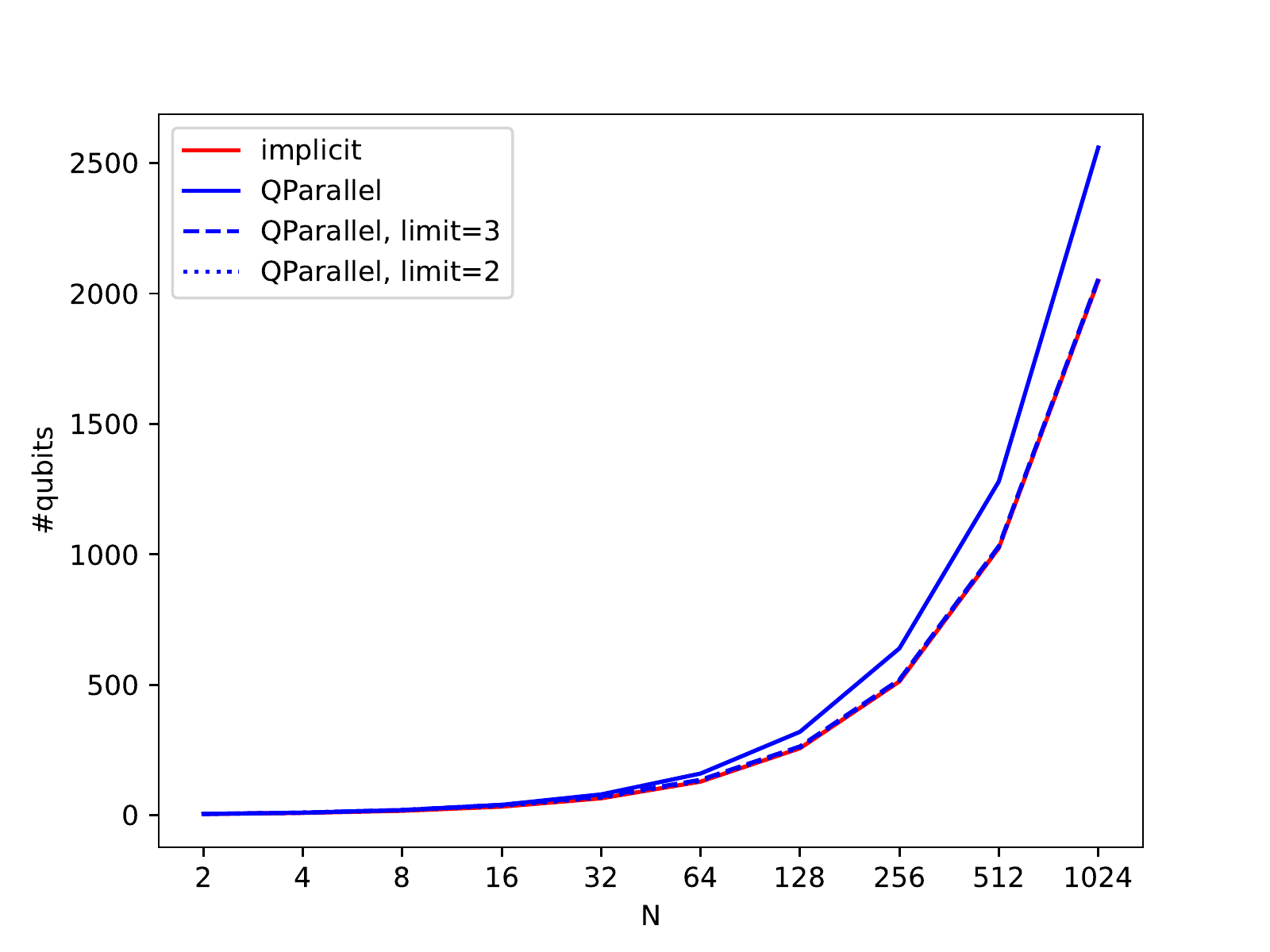}}
    \caption{Qubit count}
    \end{subfigure}
    \caption{$T$-depth and qubit count of a multi-controlled NOT gate with $N$ controls with and without using explicit parallelism (with \sy{QParallel}) in Q\#. By limiting the parallelism to a certain recursion depth, qubits can be saved at the expense of a higher $T$-depth.}
    \label{fig:ccx}
\end{figure*}

\begin{figure*}[t]
    \begin{subfigure}[t]{0.5\textwidth}
    \centering
    \resizebox{0.8\linewidth}{!}{\includegraphics{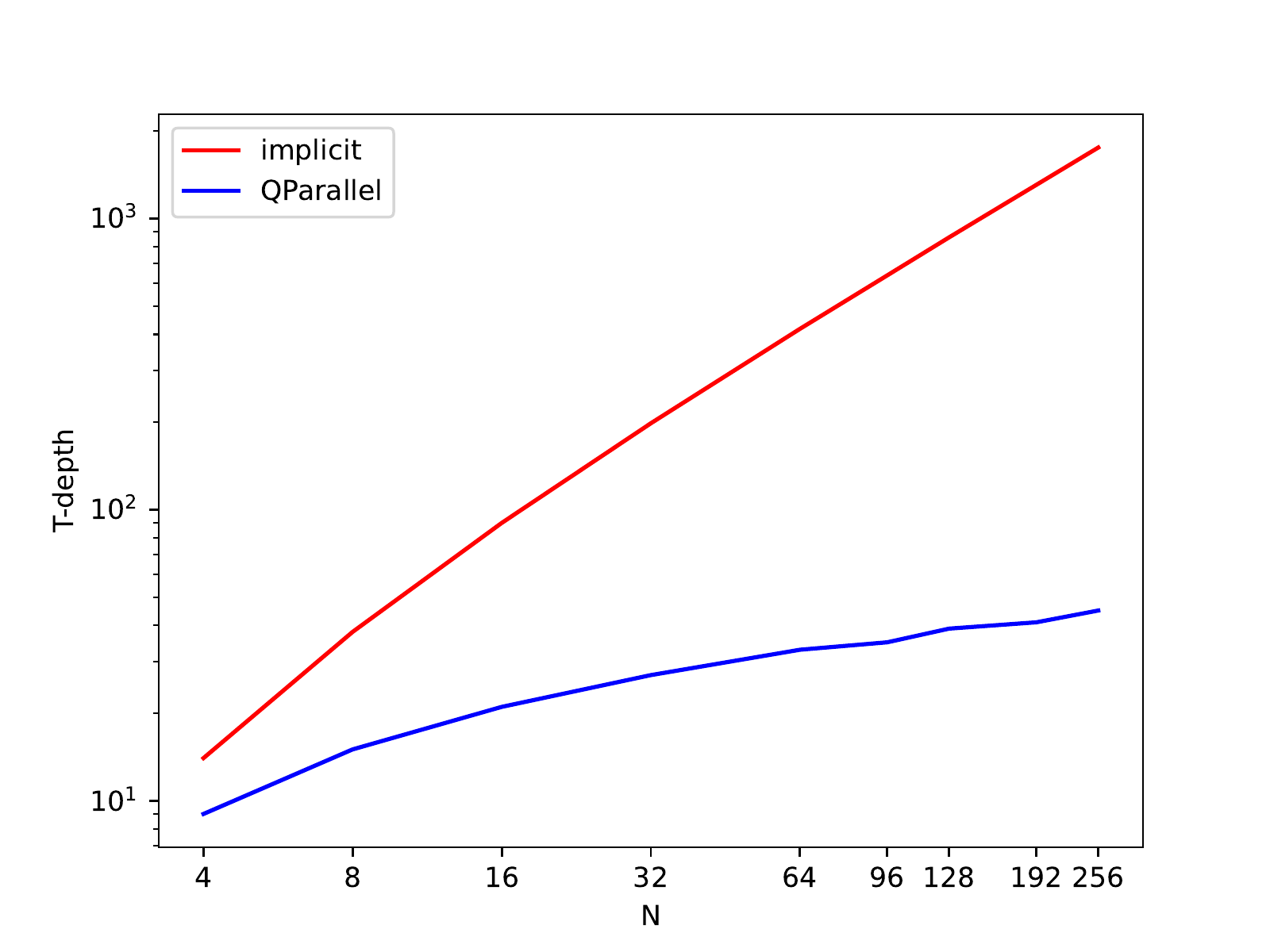}}
    \caption{$T$-depth}
    \end{subfigure}
    \begin{subfigure}[t]{0.5\textwidth}
    \centering
    \resizebox{0.8\linewidth}{!}{\includegraphics{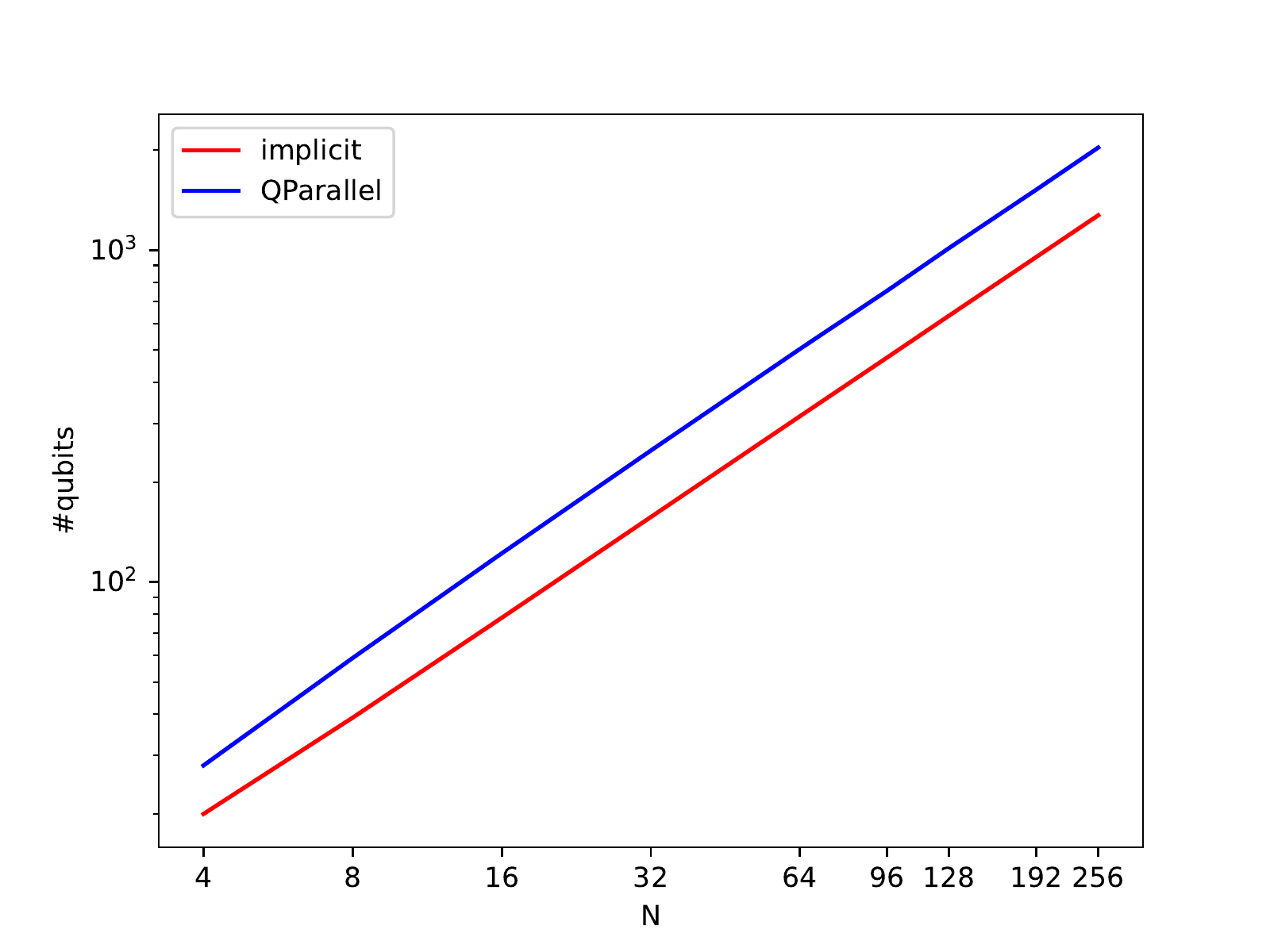}}
    \caption{Qubit count}
    \end{subfigure}
    \caption{$T$-depth and qubit count of a logarithmic-depth carry-lookahead adder~\cite{draper2004logarithmic} using implicit and explicit parallelism (with \sy{QParallel}) in Q\#. Correct behavior is observed only with explicit parallelism, which ensures that the circuit depth grows logarithmically as a function of the number of qubits $N$ used for the two numbers being added.}
    \label{fig:adder}
\end{figure*}

\begin{figure*}[t]
    \begin{subfigure}[t]{0.5\textwidth}
    \centering
    \resizebox{0.8\linewidth}{!}{\includegraphics{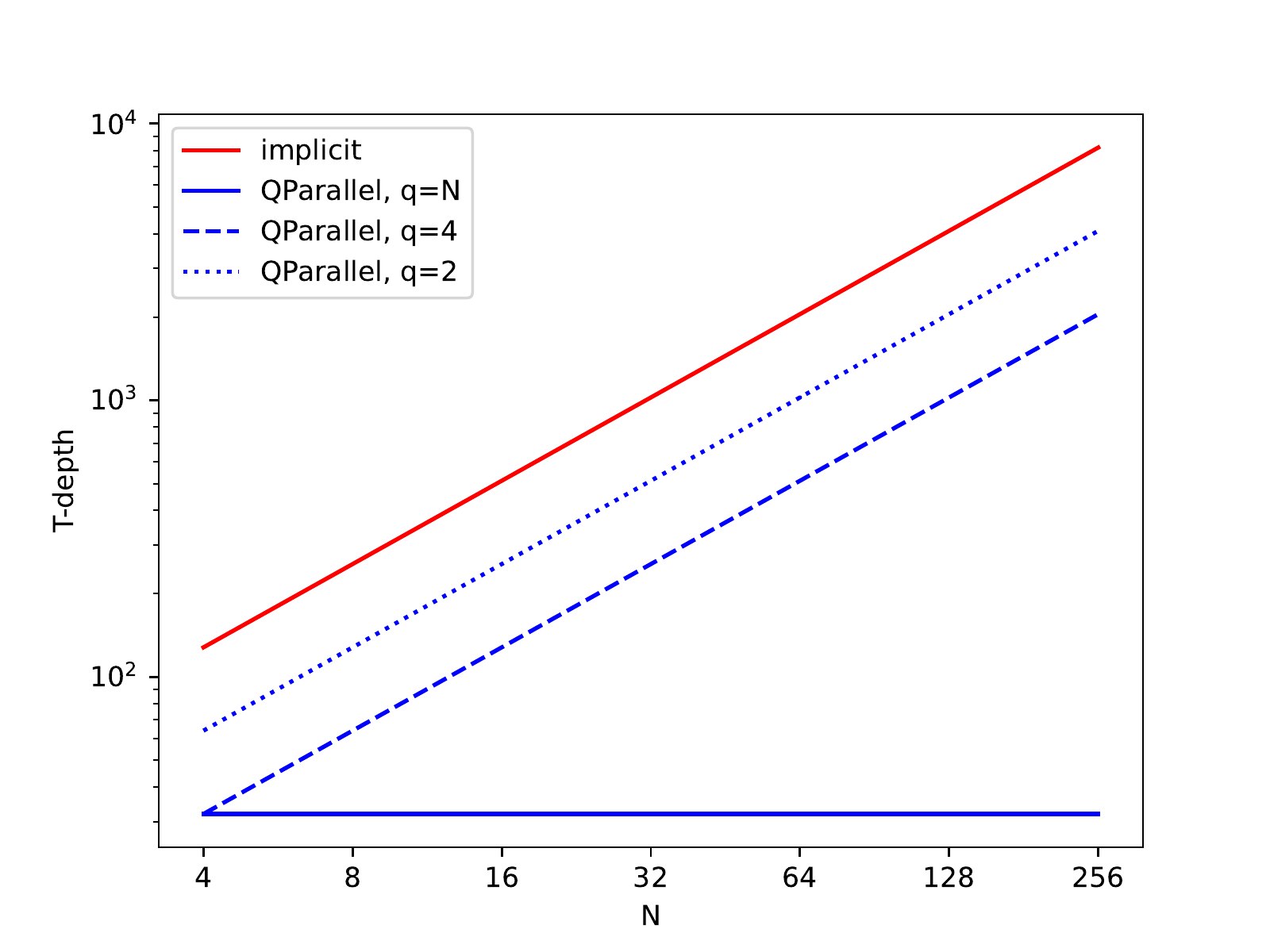}}
    \caption{$T$-depth}
    \end{subfigure}
    \begin{subfigure}[t]{0.5\textwidth}
    \centering
    \resizebox{0.8\linewidth}{!}{\includegraphics{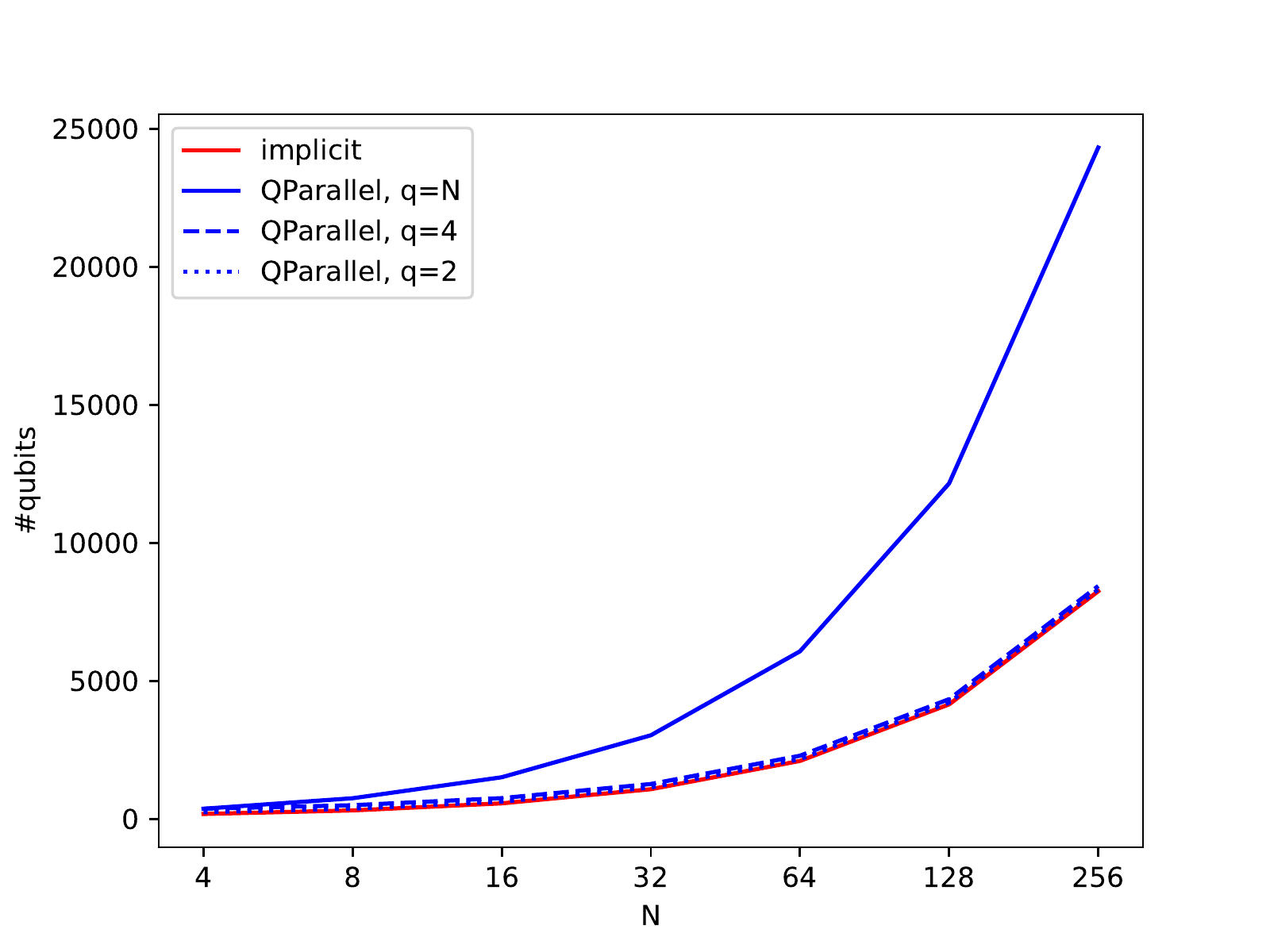}}
    \caption{Qubit count}
    \end{subfigure}
    \caption{$T$-depth and qubit count of the Givens rotation gadget implemented with $q\geq 1$ Fourier resource states and $N$ 32-bit adders from~\cite{gidney2018halving} using implicit and explicit parallelism (with \sy{QParallel}) in Q\#. Explicit parallelism ensures that the availability of $q>1$ Fourier states results in the expected parallelism.}
    \label{fig:givens}
\end{figure*}

Here, we present three examples, which were taken from practical applications in quantum chemistry and cryptography, to demonstrate how \sy{QParallel} statements facilitate exploring space-time tradeoffs in practice. For each example, we show how the explicit parallelism affects the depth and width of the resulting circuit.

Since our examples were taken from quantum algorithms featuring billions of gates at application scale~\cite{von2021quantum,haner2020improved,lee2021even,gidney2019factor}, we assume that the target quantum computer employs an error-correcting code such as the surface code to achieve the necessary gate fidelity. We thus focus on non-Clifford gates (specifically the $T$-gate) and the number of logical qubits in our analysis. We note that none of the presented space-time tradeoffs affect the number of $T$-gates, and we thus only report changes in $T$-depth.

\subsection{Multi-controlled NOT gate}
As a first introductory example, we consider the multi-controlled NOT (or X) gate, which flips the target qubit if all $n$ control qubits are 1, i.e.,
\[
    C^nX := (\mathds 1_{2^n} - \ket{\underbrace{1\cdots 1}_n}\bra{1\cdots 1})\otimes\mathds 1_2 +\ket{\underbrace{1\cdots 1}_n}\bra{1\cdots 1} \otimes X,
\]
where $X$ denotes the Pauli X gate $X=\left(\begin{smallmatrix}0&1\\1&0\end{smallmatrix}\right)$ and $\mathds 1_k$ denotes the $k\times k$-dimensional identity matrix.

It is well-known that the multi-controlled X gate can be implemented as a tree of $C^2X$ or \textit{Toffoli} gates of depth $\mathcal O(\log n)$. In turn, Toffoli gates may be implemented in terms of $CX$, $T$, $T^\dagger$, $H$, and $S$ gates in $T$-depth 1 using 4 $T$-gates and one auxiliary qubit when using measurement-based uncomputation~\cite{jones2013low}.

However, this tree-like implementation in Q\# may not result in a logarithmic-depth circuit due to the implicit reuse of the auxiliary qubit in the low-depth implementation of the Toffoli gate. Instead of the expected logarithmic depth, we observe a linear $T$-depth if we do not make use of \sy{QParallel}, see \cref{fig:ccx}. Using \sy{QParallel}, we can easily address this issue by enclosing the recursive calls in the implementation in a \sy{parallel sections} block. Moreover, we can explore space-time tradeoffs by limiting the parallelism to a certain depth in the recursion (after which the serial implementation is used). As a result, fewer qubits are required at the expense of a higher $T$-depth, see \cref{fig:ccx}.

\subsection{Log-depth adder}

As a second example, we consider a Q\# implementation of the logarithmic-depth carry-lookahead adder from~\cite{draper2004logarithmic}\rev{, extracted from an implementation to break elliptic curve cryptography on a quantum computer~\cite{haner2020improved}}.\footnote{\rev{See github.com/microsoft/QuantumEllipticCurves for the code}} Addition circuits have been heavily optimized as they are crucial components of Shor's algorithm~\cite{shor1994algorithms} and newer, highly-optimized algorithms for quantum chemistry~\cite{von2021quantum}.

Whereas adders optimized for space~\cite{takahashi2009quantum} or gate count~\cite{gidney2018halving} perform as expected when implemented in a high-level programming language such as Q\#, this is not the case for the logarithmic-depth adder from~\cite{draper2004logarithmic} due to the parallelism being captured only implicitly.

Therefore, similar to the $C^n X$ example above, the $T$-depth of the generated circuit does not grow logarithmically with the number of bits $N$ of the two numbers being added. Instead, due to qubit reuse across loop iterations, the implicitly parallel implementation becomes serial, see \cref{fig:adder}.

If we instead use \sy{parallel for} loops from \sy{QParallel} to implement the log-depth adder, we successfully address this issue. In \cref{fig:adder}, the circuit depth with explicit parallelism grows logarithmically with the bitsize $N$, as expected. \rev{This depth reduction will then affect the run-time of the application in which the adders are used, e.g., elliptic curve cryptography.}

At this point, we note that care must be taken in the implementation of the controlled version of this log-depth adder. While simply adding the control qubit(s) to each operation in the circuit will result in a correct addition circuit, its depth will again be linear due to the serial access to the control qubit(s).
Therefore, a carefully implemented version must ensure that control qubits are fanned out before entering parallel loops using the \sy{fanout(control, num)} clause provided by \sy{QParallel}'s \sy{parallel for} loop.

\subsection{Givens rotations}

As a final example, we choose a subroutine that is used in chemistry algorithms~\cite{von2021quantum} and that makes use of Fourier states as resource states to reduce the number of non-Clifford gates.

Recent quantum algorithms for quantum chemistry leverage one or multiple approximate factorizations to reduce the number of quantum gates and qubits~\cite{von2021quantum,lee2021even}. On the quantum computer, these factorizations are then undone on-the-fly to enact the desired transformation, which is used to, e.g., determine the ground state energy of the quantum system being studied.
The Givens rotation gadget, which applies a sequence of rotations where the angles are encoded in $N$ quantum registers,
is a crucial component in this transformation, and \sy{QParallel} allows programmers to investigate different space-time tradeoffs to pick the right amount of parallelism such that (1) no memory is wasted while (2) the magic state consumption rate is as close to the production rate as possible.

In our example, we assume $32$-bit angles in $N$ quantum registers and we use the adder implementation from~\cite{gidney2018halving} together with $q\geq 1$ Fourier states. Without \sy{QParallel}, the implementation does not benefit from using $q>1$ Fourier states because auxiliary qubits (in the adder implementation) are reused across the $N$ different adders.

The implementation with explicit parallelism, on the other hand, benefits from using $q>1$ Fourier states since auxiliary qubits are no longer reused for all adders. Instead, we perform the $N$ adders in chunks of $q$ fully-parallel adders, resulting in a reduction of the $T$-depth of the complete circuit by a factor of $q$, see \cref{fig:givens}.

\section{Software Stack Integration}

\begin{figure*}[t]
    \begin{subfigure}[t]{\textwidth}
    \centering
    \resizebox{0.8\linewidth}{!}{\includegraphics{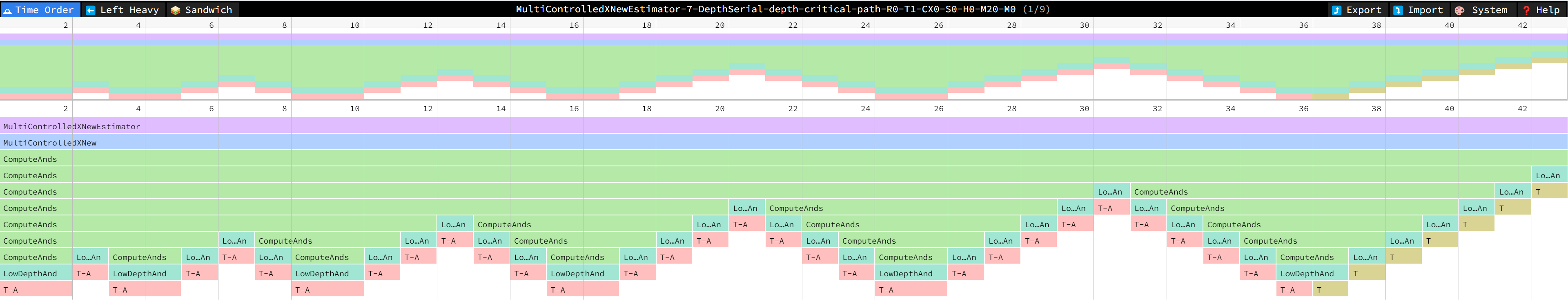}}
    \caption{Without \sy{QParallel}}
    \label{fig:speedscope_a}
    \end{subfigure}
    \begin{subfigure}[t]{\textwidth}
    \centering
    \resizebox{0.2\linewidth}{!}{\includegraphics{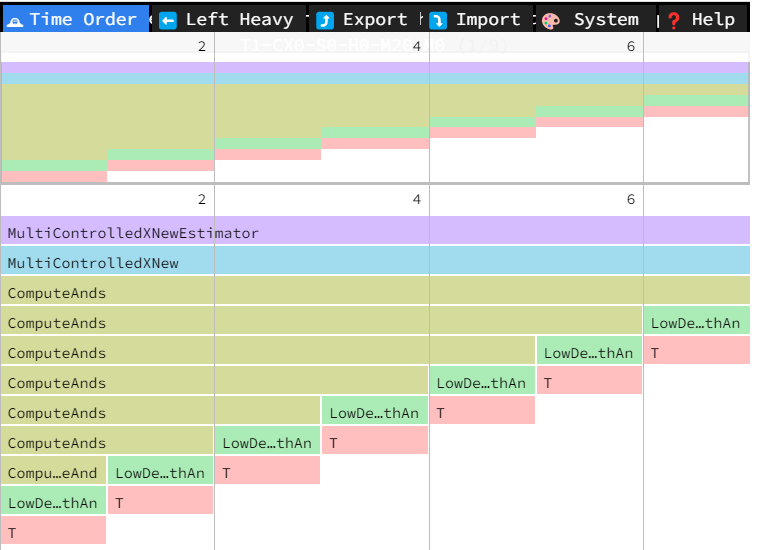}}
    \caption{With \sy{QParallel}}
    \label{fig:speedscope_b}
    \end{subfigure}
    \caption{Speedscope (\protect\url{https://www.speedscope.app}) visualization of the critical path of a multi-controlled NOT gate. Without explicit parallelism, low-depth AND gates are executed serially (due to reuse of the auxiliary qubit), whereas \sy{QParallel} makes the parallelism explicit, thus removing many operations from the critical path. The result is a highly parallel (logarithmic-depth) implementation.}
    \label{fig:speedscope}
\end{figure*}

\rev{\sy{QParallel} allows programmer, who clearly understands the ways to efficiently parallelize the algorithm to have precise control over which blocks should be parallelized and how qubits should be used in the parallel sections. This puts more control in the hands of the programmer, however, adding the burden of identifying and marking parallel sections. An optimizing compiler could, in principle, try to do some of this work implicitly, and those compiler algorithms will be a topic of the future work. But even in the presence of optimizing compiler, an ability to use precise manual controls might be useful for a programmer who wants guaranteed behavior. To ease the job of such programmer,} in addition to the \sy{QParallel} language extension, we propose a tool to help programmers identify performance-critical subroutines in a quantum program, akin to how profiling is used to find bottlenecks in classical programs.

For ease of implementation, we rely on (1) resource estimation functionality in existing software frameworks for quantum computing and (2) the flame graph visualization capabilities of speedscope.\footnote{\url{https://www.speedscope.app/}} We combine the two components by modifying the Q\# tracer such that it outputs a flame graph representation of the critical path that can be consumed by speedscope. As an example, we show flame graph visualizations of the multi-controlled NOT example with and without explicit parallelism in \cref{fig:speedscope}. \cref{fig:speedscope_a} shows that the different low-depth AND gates that get invoked in \texttt{ComputeAnds} are executed serially, in contrast to what a programmer would expect to happen (due to the implicit parallelism). Wrapping the recursive calls to \texttt{ComputeAnds} in a parallel region resolves this issue, see \cref{fig:speedscope_b}, where the total number of cycles is 7, down from the initial 43.

We have used the resulting tool extensively to optimize existing quantum programs with \sy{QParallel}, with applications ranging from cryptography to quantum chemistry. 
By visualizing the subroutines on the critical path of the quantum program, including their respective contributions to the overall runtime, programmers can quickly identify subroutines that may be removed from the critical path through the introduction of explicit parallelism. Notably, this makes it possible for programmers to add \sy{QParallel} statements in the right places even if their knowledge of the code base is very limited.

\section{Conclusions and Outlook}

Our proposed language extension, \sy{QParallel}, removes ambiguities concerning parallelism in current quantum programming languages.
Similar to classical multithreading libraries and language extensions, it facilitates investigating the tradeoffs involved in parallelizing a computation.
When combined with our tool for identifying critical paths in the code, programmers can investigate space-time tradeoffs more efficiently even if they lack detailed knowledge of the codebase. We believe that \sy{QParallel} thus leads to more rapid prototyping and optimization of large-scale quantum programs.
In turn, these optimized programs may be run on actual hardware or inform architectural decisions for future quantum hardware.

\rev{While we describe our proposed approach in terms of language extensions to Q\#, it can be applied to any quantum programming language that supports configurable quantum memory management.  We hope that our results inspire the development of current and future quantum programming languages, such as OpenQASM 3~\cite{crossOpenQasm3}, QCOR~\cite{mintz2019qcor}, and other, to include support for fine-grained control of parallel execution.}

We expect that various useful extensions to this initial version of \sy{QParallel} will be proposed and investigated in future work.
In particular, future versions may add support for the \sy{fanout} clause to parallel sections, and add more flexibility, e.g., through scheduling hints analogously to OpenMP's \sy{schedule}-clause, which allows programmers to influence how loop iterations are divided among the different threads. In \sy{QParallel}, the analogous clause could be used to influence if and how resource states and auxiliary qubits are reused.

\bibliographystyle{unsrt}
\bibliography{references}


\end{document}